\documentstyle[11pt,subeqn]{article}  
\begin{document}
\hbadness=10000
\renewcommand{\thefootnote}{\fnsymbol{footnote}}

\pagestyle{plain}
\baselineskip 18.pt

\begin{center}
{\large\bf WARD-TAKAHASHI IDENTITIES AND NOETHER'S THEOREM IN QUANTUM
FIELD THEORY}
\bigskip

Michael Danos\footnote[1]{Visiting Scholar}\\

Enrico Fermi Institute, University of Chicago\\
 Chicago, Illinois 60637\\

\end{center}

\begin{abstract}
A gap in the mathematical logic in derivations in quantum field
theory arises as consequence of variation before
quantization. To close this gap the present paper introduces
a mathematically rigorous variational calculus for operator
fields. Using quantization before variation it is
demonstrated that the so-called naive results are correct;
in particular both Noether's
theorem and the Ward-Takahashi identities retain full validity
in quantum field theory.
\end{abstract}
\bigskip

\section{Introduction}

In classical mechanics Noether's theorem plays a central role in
that, given
the Lagrangian, it allows to find the constants of the motion
without the need of actually finding solutions to the theory
\cite{1}.  As emphasized e.g., by Bjorken and Drell, Ref.~\cite{2},
Chapter 11, one must question whether the classical result,
the existence of Noether's theorem, still applies when going
over to quantum fields.   The aim of this paper is to demonstrate
that Noether's theorem can be derived in a mathematically rigorous
manner directly within the framework of quantum field theory,
and that therefore the question of its validity in quantum field theory can
be answered in the affirmative.

Elaborating on the above, in classical physics Noether's
theorem is unassailable; it arises by using variational calculus,
together with some algebraic manipulations, on the given Lagrangian
action.  All that
involves only well-understood mathematically rigorous operations.
Consequently, any results seeming to break that theorem can
immediately be declared as hiding a calculational error.  In
quantum field theory, on the other hand, in the conventional
formulation the logical structure of the derivation
is not tight in that one performs the variational
calculus, and derives the different theorems, using classical
fields.  Only afterwards one goes over to QFT: the fields are
quantized at the end, and the theorems are taken over from the
classical level \cite{2}.  Thus, the connection with the underlying
Lagrangian sis not immediate. The chain of mathematical
reasoning is broken, and one can argue that the validity
of Noether's theorem in QFT can not be taken for granted: the
theorems derived within the classical framework may not survive
the quantization procedure.  It is
the aim of the present paper to close this gap in
the mathematical logic, and hence to demonstrate the validity of
Noether's theorem in quantum field theory.

Of the mathematical manipulations, algebraic transformations
do not pose any difficulties. It is the variational calculus
which conventionally is known only for c-number fields.
As we will show, it is possible to define rigorously a generalization
of the Euler -- Lagrange variational calculus to operator fields
in such a manner that their commutation relations are maintained; hence
the thus defined variational calculus is directly valid
for operator
fields. It will be demonstrated that in terms of this calculus,
together with simple algebraic manipulations, the quantum results
turn out to be form-identical with the well-known classical results.
This way all the poweful results flowing from the application
of the variational calculus
valid in classical theory, in particular Noether's theorem,
are strictly valid also in quantum field theory.

The central problem thus is to define a variational calculus for
quantum fields, which is the subject of Section 2.  For the present
purpose one only needs to demonstrate a procedure which is
``sufficient''; it is not needed to go to the ``necessary'' stage.
To that end we shall define a variational procedure applicable to
quantum, i.e., non-commuting operator fields, which respects the
commutation relations.  We use this procedure in Section 3 to
derive the Euler -- Lagrange variational formalism.  Using that
formalism we re-derive Noether's theorem in Section 4 in order to
exhibit that indeed no mathematically ill-defined steps are
required.  In the same Section we demonstrate that for the usual
quantum fields, i.e., those needed for quantum electrodynamics, the
rigorous procedure leads precisely to the results obtained in
today's heuristic, so-called ``naive'' methods \cite{5}.  The case of PCAC
is treated in Section 5.  This way we show that all steps leading
to Noether's theorem can be carried out in a rigorous manner.
Consequently there exists no excuse for considering that
Noether's theorem may be violated in QFT, in particular, that there
may exist violations of the so-called ``naive'' Ward --
Takahashi identities.  That then has as a consequence
that some of the presently disqualified theories can be
reconsidered.

\section{Variational Derivative for Operator Fields}

We will now demonstrate that the
variational
procedure {\em can be so defined} that it respects the commutation
relations of the fields.  It
will turn out that the familiar conjecture is true that the results
of
variational calculus of \hbox{c-number} fields apply also for
operator fields upon taking care of the proper ordering of the
factors in products.

Since the basis of the variational calculus is the variational
derivative we now investigate its meaning for operator fields.  We
{\it define} it by the following limit procedure:  Let
$\varphi(x)$, $\pi(x)$, be an operator field and its canonical
conjugate, respectively, in the Heisenberg picture.  We have,
      \begin{equation}
\widetilde\varphi (x)~=~ \varphi(x)~+~ \delta\varphi(x) \label{B.1}
     \end{equation}
for the varied field and
     \begin{equation}
\widetilde\pi (x)~=~ \pi(x)~+~ \delta\pi(x) \label{B.2}
   \end{equation}
for the varied conjugate field.  We now define the variations as
     \begin{equation}
\delta\varphi (x)~=~ \varepsilon(x)~ \varphi(x) \label{B.3}
   \end{equation}
and
     \begin{equation}
\delta\pi (x)~=~ \eta (x)~ \pi(x) \label{B.4}
   \end{equation}
where $\varepsilon(x)$ and $\eta(x)$ are arbitrary \hbox{c-number}
functions which, as always, are taken to be small, i.e., to
approach zero (see below).  They can be ordinary functions or
generalized functions (for short \hbox{g-functions}).  As we will
see, these definitions suffice for a consistent variational
calculus for operator fields.   With our definitions the canonical
equal time commutation (or anti-commutation) relations for the
varied fields then are preserved in the limit.  Indeed,
   \begin{subequations}\label{B.5}
     \begin{equation}
\left[ \widetilde\varphi(x), \pi(y)\right]_{t_x=t_y} ~=~
i\delta^{(3)} ({\bf x-
y})~ [1+\varepsilon(x)] \quad , {\phantom {+\eta(y)
+\varepsilon(x)~\eta(y) }}
\label{B.5a}
   \end{equation}
or more generally
     \begin{equation}
\left[ \widetilde\varphi(x), \widetilde\pi (y)\right]_{t_x=t_y} ~=~
i\delta^{(3)} ({\bf x-y})~ [1+\varepsilon(x) +\eta(y)
+\varepsilon(x)~\eta(y)]
\label{B.5b}
    \end{equation}
\end{subequations}
and hence are preserved in the limit.  The expressions like
$\delta^{(3)} ({\bf x-y})~ \eta(y)$ in these equations which seem
to contain products of generalized functions are in fact, as shown
below, mathematically well-defined.

There is wide latitude in the choice of the \hbox{c-number}
functions.  One
such possible special choice is
    \begin{equation}
\eta(y)~=~0 \qquad \qquad \label{B.6}
\end{equation}
     \begin{equation}
\varepsilon(x)~ \Rightarrow~ \varepsilon_n(x)~=~
\lim_{\varepsilon\to 0}
~\varepsilon ~ \Delta_n^{(4)}( x-x_0) \label{B.7}
\end{equation}
\noindent
where $\varepsilon~=~0$ is excluded.  Further, $x_0$ is an
arbitrary point in space-time and $\Delta_n^{(4)} (x)$ is in
Lighthill's terminology \cite{9} a member of a set of ``good
functions'' which for $n \to \infty$ approach the
(four-dimensional) $\delta-$function:
\begin{equation}
 \delta^{(4)} (x)~=~\lim_{n\to\infty} \Delta_n^{(3)}({\bf x})
\Delta_n(t) \quad . \label{B.8}
 \end{equation}
The choice (\ref{B.6}), (\ref{B.7}) will allow the definition of
the
variational derivative at the point $x~=~({\bf x}_0,t_0)$ with
respect to the operator field $\varphi(x)$,  while maintaining the
commutation relations of the fields in the limit $\varepsilon  \to
0$.

The functional derivative with respect to the conjugate field at
the point
$y~=~({\bf y}_0,t_0)$ will arise from the choice
     \begin{equation}
\varepsilon(x) ~=~0 \qquad ~ \qquad \quad \label{B.9}
\end{equation}
     \begin{equation}
\eta (y) ~\Rightarrow~ \eta_n(y) ~=~ \lim_{\eta\to 0} ~\eta
~\Delta_n^{(4)}(y-y_0) \quad . \label{B.10}
\end{equation}
Again, the commutation relations are preserved for $\eta \to 0$.

We now give the promised demonstration of the mathematical
consistency of the terms containing $\varepsilon(x) \, \eta(y)$ in
(\ref{B.5})
when using the choices Eqs.~(\ref{B.7}), (\ref{B.10}).  They
present no
difficulties as they read
   \begin{eqnarray}
 \delta^{(3)}({\bf x-y})
[\varepsilon(x) \, \varepsilon(y)]
&=& \delta^{(3)}({\bf x-y})~
\lim_{\varepsilon\to 0} ~ \lim_{\eta\to 0} ~ [\varepsilon \, \eta]
\lim_{n\to\infty} ~ \lim_{m\to\infty} ~ \Delta_n(t-t_x)
\Delta_n({\bf x-x}_0) \qquad \quad \cr
&~& \cr
& \times &
\Delta_m^{(3)}({\bf y-y}_0)~ \Delta_m^{(3)}(t-t_y)~ \cr
&~& \cr
  & \rightarrow &
 \delta^3({\bf x-y})~\lim_{\varepsilon\to 0} ~
\lim_{\eta\to 0} ~ [\varepsilon \, \eta] \delta(t_x-t)~
\delta^{(3)}({\bf x-x}_0)~ \delta(t_y-t)~ \cr
&~& \cr
     & \rightarrow & \lim_{\varepsilon\to 0} ~ \lim_{\eta\to 0} ~
\delta(t_x-t)~ \delta^{(3)}({\bf x}-{\bf x}_0) ~\delta(t_x-t_y)~
\delta^{(3)}({\bf x}_0-{\bf y}_0) ~[\varepsilon \, \eta] \quad .
\label{B.11}
\end{eqnarray}
Thus it indeed is strictly a second-order term containing ordinary
\hbox{g-functions} with non-coincident arguments,  and not a
product of \hbox{g-functions}, as Eq.~(\ref{B.5}) seems to imply.
Of course, throughout
measure $dx$ and integration over test functions is implied.

To summarise, we have shown in this Section that it is possible to
define a
mathematically consistent variational calculus for operator fields
which
maintains the commutation relations of the field operators as a
\hbox{well-defined} mathematical limit.  The mathematical steps
involved in this procedure are all elementary.  As we will see
presently, this definition suffices for the applications needed in
QFT.

This concludes the definition of the variation of operator fields.

\section{The Euler -- Lagrange Equation}

We now investigate the functional derivative of an operator-field
Lagrangian, using to begin with the Klein-Gordon Lagrangian action
as an example:
     \begin{equation}
L_0 \{\varphi ,\partial_\mu \, \varphi\} = { {1\over2}} \int d^4x
\left[ \left(
{{\partial\varphi(x)}\over{\partial x_k}}\right)^2 - \left(
{{\partial\varphi(x)}\over{\partial t}}\right)^2 ~-~
m^2\left(\varphi(x)\right)
\right] \quad . \label{B.12}
\end{equation}
We have for the variation of the Lagrangian action
     \begin{equation}
\delta L_0 \left\{ \varphi ,\partial_\mu\varphi \right\} ~=~L_0
\left\{
\widetilde \varphi ,\partial_\mu\widetilde \varphi \right\} ~-~L_0
\left\{
\varphi ,\partial_\mu\varphi \right\} \quad . \label{B.13}
\end{equation}

In first order of the variation the mass term yields immediately
     \begin{eqnarray}
\delta_m &=& -~{m^2\over 2} \int d^4x \left[ \left(
\varphi(x)+\delta\varphi(x)
    \right) \, \left( \varphi(x)+\delta\varphi(x) \right) -
\varphi(x)\varphi(x) \right] \cr
~&~& \cr
     &=& -~{m^2\over 2} \int d^4x \left[ \varphi(x)
\delta\varphi(x)
+\delta\varphi(x) \varphi(x) \right] \label{B.14}
\end{eqnarray}
where in view of (\ref{B.3}) both terms are identical.  The space
derivative
terms are, again in first order,
  \begin{eqnarray}
 \delta_{\bf x} &=& { {1\over2}} \int d^4x \left[
{{\partial \left(\varphi(x)+\delta\varphi(x)\right)}\over{\partial
x_k}} ~
{{\partial \left(\varphi(x)+\delta\varphi(x)\right)}\over{\partial
x_k}} ~-~
{{\partial \varphi(x)} \over{\partial x_k}} ~ {{\partial
\varphi(x)} \over
{\partial x_k}} \right] \cr ~&~& \cr
&=& { {1\over2}} \int d^4x
\left\{
{{\partial \varphi(x)} \over {\partial x_k}} ~ {{\partial
\delta\varphi(x)}
\over {\partial x_k}} ~+~  {{\partial \delta \varphi(x)}
\over
{\partial x_k}} ~
          {{\partial  \varphi(x)} \over {\partial x_k}}
\right\} \cr ~&~&
\cr   &=& -{ {1\over2}} \int d^4x \left\{ \left[ {{\partial^2}
\over {\partial
x_k^2}} ~ \varphi(x) \right] ~ \delta\varphi (x) + \delta\varphi
(x) ~ \left[
{{\partial^2} \over {\partial x_k^2}} ~ \varphi(x) \right] \right\}
\quad ;
\label{B.15}
 \end{eqnarray}
again both terms are identical.  The time derivative gives
     \begin{equation}
\delta_t ~=~ { {1\over2}} \int d^4x \left\{ \left[ {{\partial^2}
\over
{\partial t^2}} ~ \varphi(x) \right] ~ \delta\varphi (x) +
\delta\varphi (x) ~
\left[ {{\partial^2} \over {\partial t^2}} ~ \varphi(x) \right]
\right\} \quad
 . \label{B.16}
 \end{equation}
Recalling the Fourier expansion of the fields
     \begin{equation}
 \varphi(x) ~=~ {1\over {(2\pi)^{3/2}}} \int {{d^3{\bf k}}   \over
{\sqrt{2\omega_k}}} ~ \left[ a_{\bf k} \, e^{ikx} + a_{\bf
k}^\dagger \, e^{-
ikx} \right] \quad \label{B.17}
 \end{equation}
we have
     \begin{equation}
{{\partial^2} \over {\partial t^2}} ~ \varphi(x) ~=~
{1\over{(2\pi)^{3/2}}}
 \int {{d^3{\bf k}} \over {\sqrt {2\omega_k}}} ~ (-\omega_k^2) \,
\left[ a_{\bf
k} \, e^{ikx} + a_{\bf k}^\dagger \, e^{-ikx} \right] \label{B.18}
\end{equation}
with
     \begin{equation}
\left[ a_{\bf k}, a_{{\bf k}'}^\dagger \right]_- ~=~ \delta^3( {\bf
k}-
{\bf k}') \quad . \label{B.19}
\end{equation}
The fields and their second time derivatives thus also commute.  To
recapitulate: in view of the definition (\ref{B.1}), (\ref{B.3}),
and
(\ref{B.7}) the fields commute with both the space-like and the
time-like
second derivatives.  This way we find
     \begin{equation}
{{\delta L} \over {\delta\varphi}} ~\delta\varphi ~=~ \left[
{\partial^2 \over
{\partial x_\mu^2}} ~\varphi(x) + m^2\varphi(x) \right]
\delta\varphi(x) \quad
. \label{B.20}
 \end{equation}
The functional derivative for operator fields thus here is
form-identical to
that for \hbox{c-number} fields.  In other words, the usual results
obtained by quantization after variation here are perfectly
legitimate, i.e., they agree with those obtained by direct
variation of quantum fields.

The case of the Dirac fields is even simpler in that no commuting
of fields and varied fields is required.  Thus we define in analogy
to (\ref{B.1}) and
(\ref{B.2})
      \begin{eqnarray}
 \delta \psi(x) &=& \varepsilon_n(x) \,
\psi (x) \label{B.21} \\
\delta \bar\psi(x) &=& \eta_n(x) \, \bar\psi (x) \label{B.22}
\end{eqnarray}
with $\varepsilon_n$ and $\eta_n$  as in (\ref{B.6}), (\ref{B.7})
or
(\ref{B.9}), (\ref{B.10}), to obtain
     \begin{eqnarray}
 \delta\bar\psi~ {{\delta L} \over {\delta\bar\psi}}  &=&
\delta\bar\psi(x) ~
(\gamma\partial +m) \, \psi (x)  \label{B.23} \\ ~&~& \cr
{{\delta L} \over
{\delta\psi}}~ \delta\psi &=& \bar \psi(x) ~ (- \gamma\partial+m)
\,
\delta\psi(x) \quad . \label{B.24}
 \end{eqnarray}
Herewith we find that the variation leading to the Euler-Lagrange
equations
immediately leads to the same results for operator fields as for
\hbox{c-number} fields.

The definitions (\ref{B.1}) through (\ref{B.5}) give an unambiguous
meaning to any form one may encounter.  Thus, the variation for the
most general case, is
     \begin{eqnarray}
\delta \, L \, \{ \varphi, \, \partial_\mu\varphi, \, \pi, \ldots
\} ~=~ L \,
\{ (\delta \varphi), \, \partial_\mu\varphi, \, \pi, \ldots \} &+&
  L \, \{
\varphi, \, \partial_\mu(\delta\varphi), \, \pi, \ldots \} \cr
~&~& \cr
&+& L \, \{ \varphi, \, \partial_\mu\varphi, \, (\delta\pi), \ldots
\} ~+~
\ldots     \quad . \label{B.25}
\end{eqnarray}

One now can freely perform integrations by parts to free the
variations from the derivative operators by the usual rules, i.e.,
as if the operator fields were \hbox{c-number} fields except that,
of course, the order of factors must be maintained.  For example,
  \begin{subequations}\label{B.26}
\begin{equation}
\int d^4x \, A(\partial_\mu\, \delta\varphi) \, B ~=~ -\int d^4x \,
(\partial_\mu\, A) \, \delta\varphi \, B ~- \int d^4x \, A \,
\delta\varphi ((\partial_\mu\, B) \label{B.26a}
\end{equation}
which, of course, can be interpreted as
\begin{equation}
\int d^4x \, \partial_\mu\, (A \, \delta\varphi \, B) ~=~ 0 \quad
.
\label{B.26b}
\end{equation}
\end{subequations}
The resulting expressions are thus precisely those one would obtain
for a
classical Lagrangian theory.  Consequently, for example,
Schwinger's
variational treatment of field operators \cite{25} this way can be
shown to be rigorously defined.  Now the commutation relations,
(\ref{B.5}), can be used to re-write the expression (\ref{B.26a})
if so desired; for example, to implement the normal ordering
prescription, or to extricate the variant as in (\ref{B.14}) or
(\ref{B.15}).

As for the interaction terms, they pose no problems as long as they
do not
contain derivatives, which, e.g., is the case for QED.  The cases
where they do contain derivatives must be individually
investigated, along the lines given in this paper.  We emphasize:
herewith we have shown that the results derived for \hbox{c-number}
fields can be derived in a rigorous manner also directly for
operator fields.

\section{Noether's Theorem}

Recalling the derivation of Noether's theorem we will see that
nothing beyond the validity of the functional derivative for
operator fields is needed. In other words, this will show that all
conservation rules derivable from Noether's theorem are strictly
valid both for \hbox{c-number} and for operator fields.

Noether's theorem concerns the consequences of the symmetries of
the Lagrangian.  That means that if the Lagrangian is not changed
as a
consequence of some transformations, be it by a transformation of
the
coordinates or a transformation of the fields, there exist
quantities
which are constants of the motion, i.e., conserved quantities; they
usually can be formulated as continuity equations.  To derive these
equations one first notes that the variation with respect to the
parameter of the transformation, say $\alpha$, vanishes:
     \begin{equation}
{{\delta L}\over{\delta\alpha}} ~=~ 0 \label{NT9.1}
\end{equation}
and then uses the Euler-Lagrange equations and some manipulations
to cast the conditions (\ref{NT9.1}) in a form of a 4-divergence,
i.e., of a differential conservation law, a continuity equation.
Here one must pay attention to the boundary conditions,
i.e., the possible existence of surface terms, as we will see
presently.  We shall first derive Noether's theorem for the case
of a
translation-invariant Lagrangian, and derive the energy -- momentum
conservation law.  In this case the variation of the fields enters
only in deriving the Euler -- Lagrange equations of motion.

Take the Lagrangian ${\cal
L}(\varphi,\partial_\mu
\varphi)$, and assume that for the replacement
     \begin{equation}
x_\mu \to  x'_\mu ~=~ x_\mu  + \delta x_\mu  \label{NT9.2}
\end{equation}
the Lagrangian is not changed.  Then we have
     \begin{equation}
\int_{\Omega '} ~{\cal L}'(x') \, d^4x' ~=~ \int_\Omega {\cal L}(x)
\, d^4x
\quad . \label{NT9.3}
\end{equation}
Hence, renaming the integration variable we obtain
     \begin{eqnarray}
0 ~=~ \delta L &=& \int_{\Omega '} ~{\cal L}' (x') \, d^4x' ~-~
\int_\Omega
~{\cal L} (x) \, d^4x \cr
~&~&\cr
&=& \int_{\Omega '} ~{\cal L}' (x) \, d^4x ~-~ \int_\Omega ~{\cal
L} (x) \,
d^4x  \quad . \label{NT9.4}
\end{eqnarray}
We now add and subtract $\int_\Omega ~{\cal L}' (x) \, d^4x$:
     \begin{equation}
\delta L ~=~ \left( \int_{\Omega'} - \int_\Omega \right) ~{\cal
L}'(x) d^4x ~+~
\int_\Omega [ {\cal L}'(x) ~-~ {\cal L}(x) ] \, d^4x \quad .
\label{NT9.5}
\end{equation}
We recognize that the first term of (\ref{NT9.5}) together with
(\ref{NT9.2})  is simply a surface term.  Herewith, up to first
order in the variation (${\cal L}' \to {\cal L}$ in the surface
term)
     \begin{equation}
\delta L ~=~ \int_\Sigma {\cal L}(x) \delta x_\mu ~ d\sigma_\mu +
\int_\Omega
\delta {\cal L}(x) \, d^4x \quad . \label{NT9.6}
\end{equation}
Consider now the second term of (\ref{NT9.6}).  The variation
$\delta {\cal L}$ here results only from the variation of the
fields
     \begin{equation}
\delta\varphi(x) ~=~ \varphi'(x) - \varphi(x) \quad , \label{NT9.7}
\end{equation}
which is treated as in Section 2. Hence
     \begin{equation}
\delta {\cal L} ~=~ {{\partial {\cal L}}\over{\partial\varphi}} ~
\delta\varphi
~+~ {{\partial {\cal L}}\over{\partial(\partial_\mu \varphi) }} ~
\partial_\mu
\delta\varphi  \quad . \label{NT9.8}
\end{equation}
Recalling the Euler -- Lagrange equation $$ {{\partial {\cal
L}}\over{\partial\varphi}} ~=~ \partial_\mu ~ {{\partial
{\cal L}}\over{\partial(\partial_\mu \varphi) }} $$
we obtain
     \begin{equation}
\delta {\cal L} ~=~ \partial_\mu ~ \left[ {{\partial {\cal L}}
\over
{\partial(\partial_\mu \varphi) }} ~ \delta\varphi \right] \quad .
\label{NT9.9}
 \end{equation}
As the last step we use the Gauss theorem
     \begin{equation}
\int_\Sigma ~f_\mu ~d\sigma_\mu ~=~ \int_\Omega ~\partial_\mu~
f_\mu d^4x
\label{NT9.10}
\end{equation}
to convert the surface integral of (\ref{NT9.6}) into a volume
integral
     \begin{equation}
\delta L ~=~ 0 ~=~ \int_\Omega ~ \partial_\mu
\left( {\cal L}(x) ~ \delta x_\mu ~+~ {{\partial {\cal L}} \over
{\partial(\partial_\mu \varphi) }} ~ \delta\varphi \right) ~d^4x
\label{NT9.11}
\end{equation}
which, owing to the arbitrariness of the variations yields
     \begin{equation}
\partial_\mu \left( {\cal L}(x) ~ \delta x_\mu ~+~ {{\partial {\cal
L}} \over
{\partial(\partial_\mu \varphi) }} ~ \delta\varphi \right) ~=~
-\partial_t \,
P_0 ~+~ \nabla \, \vec{P} ~=~ 0 \quad . \label{NT9.12}
 \end{equation}
This is the promised differential form of the continuity equation.
Indeed, no operations beyond functional derivation described in
Section 2, and algebraic manipulations, are needed.

The integral form which gives directly
the
constants of the motion is achieved by integrating
(\ref{NT9.12})
over ``all'' \hbox{3-space}, i.e., over that volume which contains
the fields, and over the time coordinate between $t_1$ and $t_2$:
 \begin{eqnarray}
0 &=& \int_{t_1}^{t_2} dt
\int d^3x~\partial_\mu \, P_\mu \cr
~&~& \cr
 &=& -\int dt \int d^3x
\left[ \partial_t \left( {\cal L}(x) \, \delta t ~+~ {{ \partial
{\cal L}}
\over { \partial (\partial_t \varphi)}} ~\delta \varphi \right)
\right] \cr
~&~& \label{NT9.13} \\
&=& - \left\{ \int d^3x \left[ {\cal L}(x) \, \delta t ~+~ {{
\partial {\cal
L}} \over { \partial \dot\varphi}} ~\delta \varphi \right]_{t_2}
~+~ \int d^3x \left[ {\cal L}(x) \, \delta t ~+~ {{ \partial {\cal
L}}
\over { \partial \dot\varphi}} ~\delta \varphi \right]_{t_1}
\right\} \quad .  \nonumber
 \end{eqnarray}

\noindent
The space-like components $\vec P$ do not survive since the fields
supposedly vanish at infinity (or, alternatively, the contribution
from the boundaries cancel when using periodic boundary
conditions).  This way we have obtained the result that
    \begin{equation}
Q ~=~ \int d^3x \left[ {\cal L}(x) ~\delta t ~+~ {{ \partial {\cal
L}}
\over { \partial \dot\varphi}} ~\delta \varphi \right]
\label{NT9.14}
\end{equation}
is time-independent, i.e., is a conserved quantity, a constant of
the motion.  Again, all operations are fully defined.

We now specify to a ``global'' translation:
     \begin{equation}
x_\mu ' ~=~ x_\mu ~+~ \varepsilon_\mu \quad ; \quad \delta x_\mu
~=~
\varepsilon_\mu \label{NT9.15}
\end{equation}
and
     \begin{equation}
\varphi'(x') ~=~ \varphi(x) \label{NT9.16}
\end{equation}
which, for example, for a plane wave would read with (\ref{NT9.7}),
in first
order $$ \varphi '(x') ~=~ (1 - ik_\mu \varepsilon_\mu) ~
\varphi(x')
~\simeq~ e^{-ik_\mu \varepsilon_\mu} ~ e^{ik_\mu x'_\mu} ~=~
e^{ikx}
~=~ \varphi(x) \quad . $$ Now we manipulate (\ref{NT9.16}) as
     \begin{eqnarray}
0 &=& \varphi '(x') ~-~ \varphi(x) ~=~ \varphi'(x') ~-~ \varphi(x')
~+~
\varphi(x') ~-~ \varphi(x) \cr
&~&\cr
     &=& \delta \varphi(x') ~+~ \varepsilon_\nu \, \partial_\nu ~
\varphi(x)
\label{NT9.17}
\end{eqnarray}
which, inserted in (\ref{NT9.14}) leads to
     \begin{eqnarray}
P_0 &=& \int d^3x \left[ - i {\cal L}(x) ~ \varepsilon_4 ~-~ {{
\partial {\cal
L}} \over { \partial \dot\varphi}} ~\varepsilon_\nu \, \partial_\nu
\varphi
\right] \cr
&~&\cr
&=& \int d^3x \, \varepsilon_\nu \left[ - i\, \delta_{\nu 4}~ {\cal
L}(x) ~-~
{{ \partial {\cal L}} \over { \partial \dot\varphi}} ~\partial_\nu
\varphi
\right] \label{NT9.18}
\end{eqnarray}
as the conserved quantity.  Returning to (\ref{NT9.12}) we re-write
it for our case using (\ref{NT9.17}) as
     \begin{eqnarray}
0 &=& \partial_\mu \left( {\cal L}(x) ~ \varepsilon_\mu ~-~
{{ \partial {\cal L}} \over { \partial (\partial_\mu \varphi) }}
~\varepsilon_\nu \, \partial_\nu \varphi \right) ~=~ \partial_\mu
\left( {\cal
L} \, \delta_{\mu\nu} ~-~ {{ \partial {\cal L}} \over { \partial
(\partial_\mu
\varphi) }} ~ \partial_\nu \varphi \right) ~ \varepsilon_\nu \cr
&~&\cr
&\equiv& \partial_\mu ~ {\cal T}_{\mu\nu} ~ \varepsilon_\nu \quad
.
\label{NT9.19}
\end{eqnarray}
Owing to the arbitrariness of $\varepsilon_\nu$ there must hold
     \begin{equation}
\partial_\mu ~{\cal T}_{\mu\nu} ~=~ 0 \label{NT9.20}
\end{equation}
with
     \begin{equation}
{\cal T}_{\mu\nu} ~=~ {\cal L} \, \delta_{\mu\nu} ~-~ {{ \partial
{\cal L}}
\over { \partial (\partial_\mu \varphi) }} ~ \partial_\nu \varphi
\label{NT9.21}
\end{equation}
and, comparing with (\ref{NT9.18})
     \begin{equation}
P_\mu ~=~ -i \int {\cal T}_{4\mu} ~ d^3x \quad . \label{NT9.22}
\end{equation}

We recapitulate:  using the validity of (\ref{NT9.15}) we have
derived the conservation law (\ref{NT9.21}), and, more specifically
(\ref{NT9.18}), needing no mathematical operations beyond the
functional derivative and algebraic manipulations.  If the symmetry
of (\ref{NT9.15}) is the only symmetry of the Lagrangian then the
above conservation laws are the only ones guaranteed by the
Lagrangian.  Since, in order to be useful in the description of
Nature, a theory must guarantee energy-momentum conservation, it
suggests itself to identify $P_\mu$, (\ref{NT9.22}), with the
energy -- momentum \hbox{four-vector}, and ${\cal T}_{\mu\nu}$,
(\ref{NT9.19}), with the stress tensor.  Once the identification of
${\cal T}_{\mu\nu}$ as the stress tensor has been accepted, it must
be demanded to be valid in any and every theory, now in the precise
form:  the conservation law arising from the translation invariance
(if it exists in the considered Lagrangian) concerns and yields the
energy -- momentum conservation law of the theory.  And fully
generally:  if the Lagrangian has some symmetry leading to a
conservation law as in (\ref{NT9.14}), then, if a solution seems to
violate that law, the calculation must contain an error.

Because of its importance we derive one more conservation law,
which will
provide an example of the analysis concerning ``internal''
symmetries.

We consider the Lagrangian
     \begin{eqnarray}
{\cal L} &=& - \bar\psi
\left[ \gamma_\mu \left( \partial_\mu - i \, e \, A_\mu \right) +
m \right]
\psi + {\cal L}(F_{\mu\nu}) \cr
&~&\cr
&=& - \bar\psi \, \gamma_\mu \partial_\mu \, \psi + i \, e \,
\bar\psi \,
\gamma_\mu \, A_\mu \, \psi - \bar\psi \, \psi \, m + {\cal
L}(F_{\mu\nu})
\quad .   \label{NT9.23}
 \end{eqnarray}
Here $\psi(x)$ is a Dirac spinor field, and $A(x)$ stands for the
electromagnetic vector potential.  Hence this Lagrangian is said to
describe
``spinor electrodynamics.''

Since $\psi$, in contrast to $\varphi$ above, is complex, and since
in
observables a change of the phase is irrelevant, (\ref{NT9.23})
should be
invariant under the transformation
\begin{subequations}\label{NT9.24}
     \begin{equation}
\psi \to \psi ' ~=~ e^{i\alpha} ~ \psi \quad . \label{NT9.24a}
\end{equation}
   \begin{equation}
\bar\psi \to \bar\psi ' ~=~ e^{-i\alpha} ~ \bar\psi \quad .
\label{NT9.24b}
\end{equation}
\end{subequations}
We thus require
     \begin{eqnarray}
\delta L ~=~ 0 &=& - \int \left[ \delta \bar\psi ~ {{ \partial
{\cal L}} \over
{ \partial \bar\psi }} ~+~ {{ \partial {\cal L}} \over { \partial
\psi }} ~
\delta\psi ~+~ \partial_\mu (\delta \bar\psi) ~ {{ \partial {\cal
L}} \over
{\partial (\partial_\mu \bar\psi) }} ~+~ {{ \partial {\cal L}}
\over  {\partial
(\partial_\mu \psi) }} ~ \partial_\mu (\delta\psi) \right. \cr
~&~& \cr
&~& \quad  ~+~ \left. {{ \partial {\cal L}} \over { \partial A }}
~\delta A ~+~ {{ \partial {\cal L}} \over  {\partial (\partial_\mu
A) }} ~
\partial_\mu (\delta A)\right] d^4x \quad .
\label{NT9.25}
 \end{eqnarray}
Taking $\alpha$ to be infinitesimal we have from (\ref{NT9.24})
     \begin{eqnarray}
{{ \delta\bar\psi} \over {\delta\alpha}} &=& -i \, \bar\psi \cr
~&~&\cr
{{ \delta \psi} \over {\delta\alpha}} &=& i \, \psi \cr
 ~&~&\cr
\partial_\mu \, \delta\psi &=& (i \, \partial_\mu \alpha)~ \psi ~+~
i \, \alpha
\, \partial_\mu \psi \cr
  ~&~&\cr
\partial_\mu ~ \, {{ \delta \psi} \over {\delta\alpha}} &=& i \, {{
\delta
\psi} \over {\delta\alpha}} \label{NT9.26}
\end{eqnarray}
and thus
     \begin{eqnarray}
\delta L = 0 &=& - \int \left\{ - i \, \alpha \bar\psi
\left[ \gamma_\mu (\partial_\mu - i \, e A_\mu ) + m \right]   \psi
~+~
\bar\psi \, \gamma_\mu \left[ ( i \partial_\mu \alpha )  \psi + i
\alpha
\partial_\mu \psi \right] \right. \cr
  ~&~&\cr
&~& \quad ~\quad \left. - i \, e \bar\psi \gamma_\mu \psi \delta
A_\mu \right\}
\quad . \label{NT9.27}
 \end{eqnarray}
Since $\alpha$ is an arbitrary function of $x$, (\ref{NT9.27})
imposes the
condition
     \begin{equation}
e \delta A_\mu ~=~ \partial_\mu \alpha \label{NT9.28}
\end{equation}
for the Lagrangian (\ref{NT9.23}) to be invariant under the
transformation
(\ref{NT9.24}). Owing to the antisymmetry of $F_{\mu\nu}$ the last
term of
(\ref{NT9.23}) does not yield a contribution.

We now apply the Euler-Lagrange equation
\begin{subequations}
     \begin{eqnarray}
{{\delta L} \over {\delta \bar\psi}} &=& 0 ~\Rightarrow~ {{\partial
{\cal L}}
\over {\partial \bar\psi}}      ~=~ 0 \label{NT9.29a} \\
~&~&\cr
{{\delta L} \over {\delta \psi}} &=& 0 ~\Rightarrow~ {{\partial
{\cal L}} \over
{\partial \psi}} ~-~ {\partial \over {\partial_\mu}}~ {{ \partial
{\cal L}}
\over { \partial (\partial_\mu \psi) }} ~=~ 0 \label{NT9.29b}
\end{eqnarray}
to obtain
     \begin{equation}
\left( {{\partial {\cal L}} \over {\partial \psi}} \right) ~
\delta\psi ~=~
\left( {\partial \over {\partial_\mu}}~ {{ \partial {\cal L}} \over
{ \partial
(\partial_\mu \psi) }} \right) ~\delta\psi \quad . \label{NT9.29c}
\end{equation}
\end{subequations}

\noindent
Inserting this in (\ref{NT9.25}) we find
     \begin{eqnarray}
\delta L &=& 0 ~\Rightarrow~ \left[ \partial_\mu \left( {{\partial
{\cal L}}
\over {\partial (\partial_\mu \psi) }} \right) ~ \delta\psi ~+~
{{\partial
{\cal L}} \over {\partial (\partial_\mu \psi) }} ~ \partial_\mu~
\delta\psi
\right] \cr
 ~&~&\cr
&=& \partial_\mu \left[ {{\partial {\cal L}} \over {\partial
(\partial_\mu
\psi) }} ~ \delta\psi \right] ~=~ \partial_\mu \left[ {{\partial
{\cal L}}
\over {\partial (\partial_\mu \psi) }} ~ i \, \alpha \, \psi
\right] \cr
~&~&\cr
 &\Rightarrow & i~ {\partial \over {\partial_\mu}} ~ \left[ {{
\partial {\cal
L}} \over { \partial (\partial_\mu \psi) }} ~ \psi \right] ~=~ 0
\label{NT9.30}
 \end{eqnarray}
the last step again owing to the arbitrariness of $\alpha$.  This
yields, in
view of (\ref{NT9.23})
    \begin{equation}
j_\mu ~=~ i~ {{ \partial {\cal L}} \over { \partial (\partial_\mu
\psi) }}~
\psi ~=~ i \, \bar\psi \, \gamma_\mu \psi \label{NT9.31}
\end{equation}
as expected.  And, of course, the Lagrangian, being translation
invariant,
leads to a stress tensor analogous to (\ref{NT9.21}).

Again, only the functional derivative and some algebraic
manipulations are
needed as mathematical operations in the above derivations.

\section{Axial Currents}

As the last example we investigate the question of axial current
anomalies \cite{3},\cite{4}.  As is well-known, they violate the
Ward -- Takahashi
identities which arise
directly as consequences of the Euler -- Lagrange equations.
Indeed, writing out the Euler -- Lagrange equations (\ref{NT9.29a}),
(\ref{NT9.29b}), which arise from the Lagrangian
Eq.~(\ref{NT9.23}), we have:
\begin{subequations}
     \begin{eqnarray}
0 &=&     {{\partial }
\over {\partial \bar\psi}}   \left( - \bar\psi
\left[ \gamma_\mu \left( \partial_\mu - i \, e \, A_\mu \right) +
m \right]
\psi + {\cal L}(F_{\mu\nu})\right) \cr
&~&\cr
&=& \gamma_\mu \partial_\mu \, \psi + i \, e \, \gamma_\mu \, A_\mu
\, \psi -
\psi \, m \quad ,  \label{5.1a}
\end{eqnarray}
and
\begin{eqnarray}
0 &=& \left( {{\partial } \over
{\partial \psi}} ~-~ {\partial \over {\partial_\mu}}~ {{ \partial
}
\over { \partial (\partial_\mu \psi) }}\right) \left(  - \bar\psi
\left[ \gamma_\mu \left( \partial_\mu - i \, e \, A_\mu \right) +
m \right]
\psi + {\cal L}(F_{\mu\nu})\right) \cr
&~&\cr
 &=& - \partial_\mu \,\bar\psi \, \gamma_\mu  + i \, e \, \bar\psi
\,
\gamma_\mu \, A_\mu  - \bar\psi \, m   \quad .  \label{5.1b}
\end{eqnarray}
\end{subequations}
Multiplying (\ref{5.1a}) on the left by $\bar\psi \, \gamma_5$ and
(\ref{5.1b})
on the right  by $\gamma_5\, \psi$ and adding these equations we
obtain
\begin{equation}
 \partial_\mu  \bar\psi \,\gamma_\mu \gamma_5\, \psi ~=~ 2im
\bar\psi \,
\gamma_5\, \psi\       \label{5.2}
\end{equation}
which is the basis of the usual, so-called ``naive''  axial-vector
Ward -- Takahashi identity. Again, after the variational derivative
only strictly rigorous mathematical operations are needed in the
derivation. For the
discussion of the mathematical inaccuracy responsible for the
anomalous
breaking of the Ward -- Takahashi identity Eq.~(\ref{5.2}) see ref
\cite{6}.

This way, in all the above examples, no mathematically
\hbox{ill-defined},
questionable operations are required in the derivations.

\section{Conclusions}

All the results obtained in the examples of this paper, from the
definition of the variational calculus for quantum fields, up to
the derivation of Noether's theorem, were obtained without the use
of any \hbox{ill-defined} mathematical steps or concepts.  Thus
there is no need to check whether the
conservation laws obtained on the \hbox{c-number} level from
Noether's theorem ``are consistent with the commutation relations''
\cite{2}.  The previous gap in the mathematical logic has been
closed since the variational calculus has been constructed
precisely so as to be applicable directly to operator fields, i.e.,
to ensure that the commutation relations are maintained in the
variational procedure.  And, as we have shown, the results are
those one would have obtained for \hbox{c-number} fields.

This way, all results obtained in the so-called ``naive'' manner,
i.e.,
performing ``quantization after variation'', remain valid; in
particular, all ``naive'' Ward -- Takahashi identities retain
validity in QFT.  Thus, Noether's theorem is fully valid
in quantum field theory.

\end{document}